\documentclass[a4paper,11pt]{article}
\usepackage{pos}
\usepackage{subcaption}
\usepackage{dsfont}
\usepackage{bm}
\usepackage[braket,qm]{qcircuit}
\usepackage{color}

\renewcommand{\sp}{\ensuremath{\sigma^+}}
\newcommand{\sm}{\ensuremath{\sigma^-}}
\newcommand{\sx}{\ensuremath{\sigma^x}}
\newcommand{\sy}{\ensuremath{\sigma^y}}
\newcommand{\sz}{\ensuremath{\sigma^z}}
\renewcommand{\vec}{\bm}

\title{Exploring the phase structure of the multi-flavor Schwinger model with quantum computing}
\ShortTitle{Phase structure of the multi-flavor Schwinger model with quantum computing}

\author[a,b]{Lena Funcke}
\author[c]{Tobias Hartung}
\author[d]{Karl Jansen}
\author*[d,e]{Stefan Kühn}
\author[f]{Marc-Oliver Pleinert}
\author[f]{Stephan Schuster}
\author[f]{Joachim von Zanthier}

\affiliation[a]{Transdisciplinary Research Area ``Building Blocks of Matter and Fundamental Interactions'' (TRA Matter) and Helmholtz Institute for Radiation and Nuclear Physics (HISKP), University of Bonn, Nußallee 14-16, 53115 Bonn, Germany}
\affiliation[b]{Center for Theoretical Physics, Co-Design Center for Quantum Advantage, and NSF AI Institute for Artificial Intelligence and Fundamental Interactions, Massachusetts Institute of Technology, 77 Massachusetts Avenue, Cambridge, MA 02139, USA}
\affiliation[c]{Northeastern University - London, Devon House, St Katharine Docks, London E1W 1LP, UK}
\affiliation[d]{Deutsches Elektronen-Synchrotron DESY, Platanenallee 6, 15738 Zeuthen, Germany}
\affiliation[e]{Computation-Based Science and Technology Research Center, The Cyprus Institute, 20 Kavafi Street, 2121 Nicosia, Cyprus}
\affiliation[f]{Physics Department, Friedrich-Alexander Universität Erlangen-Nürnberg (FAU), Quantum Optics and Quantum Information, Staudtstr. 1, 91058 Erlangen, Germany}

\emailAdd{lfuncke@uni-bonn.de}
\emailAdd{tobias.hartung@desy.de}
\emailAdd{karl.jansen@desy.de}
\emailAdd{stefan.kuehn@desy.de}
\emailAdd{marc.pleinert@fau.de}
\emailAdd{stephan.schuster@fau.de}
\emailAdd{joachim.vonzanthier@fau.de}

\abstract{We propose a variational quantum eigensolver suitable for exploring the phase structure of the multi-flavor Schwinger model in the presence of a chemical potential. The parametric ansatz circuit we design is capable of incorporating the symmetries of the model, present in certain parameter regimes, which allows for reducing the number of variational parameters substantially. Moreover, the ansatz circuit can be implementated on both measurement-based and circuit-based quantum hardware. We numerically demonstrate that our ansatz circuit is able to capture the phase structure of the model and allows for faithfully approximating the ground state. Our results show that our approach is suitable for current intermediate-scale quantum hardware and can be readily implemented on existing quantum devices.\\

Preprint number: MIT-CTP/5479}

\FullConference{%
  The 39th International Symposium on Lattice Field Theory (Lattice2022),\\
  8-13 August, 2022 \\
  Bonn, Germany 
}

\begin{document}
\maketitle

\section{Introduction}
The advent of quantum technology during recent years provides an alternative avenue towards computationally tackling lattice field theories. Using the Hamiltonian lattice formulation, they can be directly simulated on quantum devices, thus evading the infamous sign problem that hinders conventional classical Monte Carlo (MC) simulations in certain parameter regimes. Prominent examples are the presence of a topological $\theta$-term, finite baryon density, or out-of-equilibrium dynamics, situations which are largely inaccessible with MC methods. This major promise of quantum computing has already been demonstrated successfully in several proof-of-principle experiments~\cite{Martinez2016,Klco2018,Atas2021,Mazzola2021,Ciavarella2021}, thus showing the potential of quantum computers to explore parameter regimes that are out of reach with the MC approach.

Current quantum devices are of intermediate scale and still suffer from a considerable level of noise. Hence, in order to utilize their potential, appropriate algorithms in combination with circuit optimization and error mitigation techniques are required~\cite{Endo2018,Funcke2020,Funcke2020a,Funcke2021a}. A particularly well-suited approach for these kinds of devices is the variational quantum eigensolver (VQE)~\cite{Peruzzo2014,McClean2016}. This hybrid quantum-classical algorithm allows for approximating the ground state of a given Hamiltonian using a parametric quantum circuit as a variational ansatz. A crucial aspect for the performance of the VQE is the choice of the variational ansatz. On the one hand, it should be sufficiently simple to be implemented on current noisy, intermediate-scale quantum devices. On the other hand, it should be expressive enough to capture the relevant physics of the model. Moreover, it is desirable to incorporate the relevant symmetries of the underlying Hamiltonian.

In these proceedings, we develop a suitable ansatz for a VQE to study the Schwinger model with three fermion flavors in the presence of a chemical potential, which is a regime where conventional MC methods suffer in general from the sign problem. We demonstrate that our ansatz is able to capture the relevant physics of the model at a low circuit depth and how to incorporate the symmetries of the model. Interestingly, our ansatz is not only suitable for gate-based digital quantum hardware but can also readily be translated to measurement-based quantum computers~\cite{Raussendorf2001,Nielsen2006,Ferguson2021}.

These proceedings are organized as follows. In Sec.~\ref{sec:model} we briefly introduce the Hamiltonian lattice formulation of the multi-flavor Schwinger model before presenting our ansatz circuit in Sec.~\ref{sec:ansatz}. Finally, we present our results on the performance of the ansatz in various parameter regimes in Sec.~\ref{sec:results} before concluding in Sec.~\ref{sec:conclusion}.

\section{The Schwinger model on the lattice\label{sec:model}}
In order to run the VQE, we need a Hamiltonian lattice formulation of the model. The Schwinger Hamiltonian for $F$ fermion flavors on a lattice with spacing $a$ and $N$ sites reads~\cite{Banuls2016a}
\begin{align}
  \begin{aligned}
    H = &-\frac{i}{2a}\sum_{n=0}^{N-2}\sum_{f=0}^{F-1}\left(\phi^\dagger_{n,f}e^{i\theta_n}\phi_{n+1,f}-\mathrm{h.c.}\right)+\sum_{n=0}^{N-1}\sum_{f=0}^{F-1}\left(m_f(-1)^n +\kappa_f \right)\phi^\dagger_{n,f}\phi_{n,f}+ \frac{g^2 a}{2}\sum_{n=0}^{N-2} L_n^2,
  \end{aligned}
  \label{eq:hamiltonian}
\end{align}
where we have used staggered fermions. In the expression above, $\phi_{n,f}$ ($\phi_{n,f}^\dagger$) describes a single-component matter field annihilating (creating) a fermion of flavor $f$ at site $n$, and the operators $L_n$ and $e^{i\theta_n}$ act on the links in between two matter sites $n$ and $n+1$. $L_n$ corresponds to the electric field on link $n$, and the operator $\theta_n$ is its canonical conjugate, $[\theta_n,L_{n'}] = i\delta_{nn'}$. Hence, $e^{i\theta_n}$ acts as a lowering operator in the eigenbasis of the electric field $L_n$. The parameters $m_f$ and $\kappa_f$ correspond to the bare mass and the bare chemical potential for flavor $f$, while $g$ denotes the bare coupling. In addition, the physical states $\ket{\psi}$ of the Hamiltonian in Eq.~\eqref{eq:hamiltonian} have to fulfill Gauss law, $\forall n:\ G_n\ket{\psi} = q_n\ket{\psi}$, where
\begin{align}
 G_n = L_n  - L_{n-1} - Q_n
 \label{eq:gauss_law}
\end{align}
are the generators for time-independent gauge transformations and $Q_n = \sum_{f=0}^{F-1}\phi_{n,f}^\dagger\phi_{n,f}-\frac{F}{2}(1-(-1)^n)$ is the staggered charge. The integer values $q_n$ correspond to static external charges and for the rest of the paper we choose to work in the sector of vanishing external charges, $\forall n:\ q_n=0$.

For open boundary conditions, Eq.~\eqref{eq:gauss_law} allows us to recursively reconstruct the electric field values from the matter content of the sites after fixing the value $l_{-1}$ of the electric field on the left boundary, $L_n = l_{-1}+ \sum_{k=0}^n Q_k$. Choosing $l_{-1}=0$, inserting the expression for $L_n$ into Eq.~\eqref{eq:hamiltonian} and applying a residual gauge transformation~\cite{Hamer1997,Sala2018}, we obtain
\begin{align}
\begin{aligned}
 W = -ix\sum_{n=0}^{N-2}\sum_{f=0}^{F-1}\left(\phi^\dagger_{n,f}\phi_{n+1,f}-\mathrm{h.c.}\right) +\sum_{n=0}^{N-1}\sum_{f=0}^{F-1}\left(\mu_f(-1)^n +\nu_f \right)\phi^\dagger_{n,f}\phi_{n,f}  + \sum_{n=0}^{N-2} \left( \sum_{k=0}^n Q_k \right)^2,
\end{aligned}
\label{eq:hamiltonian_dimensionless}
\end{align}
where $x=1/(ag)^2$, $\mu_f = 2\sqrt{x}m_f/g$, and $\nu_f = 2\sqrt{x}\kappa_f/g$. Equation~\eqref{eq:hamiltonian_dimensionless} shows that all notion of the gauge field is gone and we obtain a formulation directly on the gauge-invariant subspace at the expense of creating long-range interactions. 

Our goal is to determine the phase structure of the Hamiltonian~\eqref{eq:hamiltonian_dimensionless} using the VQE. To this end, we translate the fermionic degrees of freedom into spins using a Jordan Wigner transformation~\cite{Hamer1997,Martinez2016}. The different types of terms in the Hamiltonian are mapped according to
\begin{align}
\phi^\dagger_{n,f}\phi_{n+1,f} \to \sp_{n,f}(i\sz_{n,f})\dots(i\sz_{n+1,f-1})\sm_{n+1,f},\quad\quad
\phi^\dagger_{n,f}\phi_{n,f} \to \frac{1}{2}\left(\sz_{n,f} + \mathds{1}\right),
\end{align}
where $\sigma^\pm = (\sx \pm i \sy)/2$ and $\sx$, $\sy$ and $\sz$ are the usual Pauli matrices. Equation~\eqref{eq:hamiltonian_dimensionless} is thus transformed into a spin Hamiltonian for $NF$ spins. This form allows for measuring the expectation value of the Hamiltonian on a quantum device by arranging the different Pauli terms into commuting groups and measuring them individually\footnote{Note that in general there is no unique choice for arranging the Pauli terms into commuting groups. While different groupings do not affect the expectation value, the variances are in general not the same~\cite{Funcke2020}.}. Note that, for the special choice $\nu_f=-\nu_{F-1-f}$ and $\mu_f=\mu_{F-1-f}$, the spin Hamiltonian is invariant under flipping all spins followed by a spatial reflection around the center of the system, i.e.\ the transformation $\sigma^{a}_j \to (\sx\sigma^{a}\sx)_{NF-1-j}$ for $a\in\{x,y,z\}$. In addition, the Hamiltonian conserves the total charge and we restrict ourselves to the the sector of vanishing total charge, $\sum_{n=0}^{N-1} Q_n = 0$.

In the following, we focus on the case of three fermion flavors, for which the phase structure has been determined analytically for the special case $\mu_f=0$. It was found that the physics only depends on the differences $\nu_0-\nu_1$ and $\nu_2-\nu_1$, and the model goes through a series of first-order quantum phase transitions~\cite{Lohmayer2013}. The different phases are characterized by $(\Delta N_0, \Delta N_2)$ where $\Delta N_f$ is the expectation value of $\sum_{n=0}^{N-1}(\phi^\dagger_{n,f}\phi_{n,f} - \phi^\dagger_{n,1}\phi_{n,1})$. Since the Hamiltonian conserves the total particle number as well as the individual particle numbers for each flavor, $\Delta N_f$ can only take integer values. Thus, an abrupt jump in $(\Delta N_0, \Delta N_2)$ signals the onset of a first-order phase transition.

\section{VQE ansatz circuit\label{sec:ansatz}}
For the investigation of the phase structure of the Hamiltonian in Eq.~\eqref{eq:hamiltonian_dimensionless}, we use the VQE to approximate the ground state of the model. Subsequently, we extract the observable $\Delta N_f$ from measuring the resulting state in the $Z$-basis. To obtain the ground state, the VQE iteratively minimizes the cost function
\begin{align}
  C(\vec{\theta}_k) = \melem{\psi(\vec{\theta}_k)}{W}{\psi(\vec{\theta}_k)}
\end{align}
using the quantum device to efficiently evaluate the energy expectation value for a given set of variational parameters $\vec{\theta}_k\in\mathds{R}^p$. Based on the measurement outcome, a classical optimization algorithm is used to determine a new set of parameters $\vec{\theta}_{k+1}$ to decrease the cost function. Running the hybrid quantum-classical feedback loop until convergence, the solution $\ket{\psi(\vec{\theta}^*)}$ encodes a good approximation for the ground state of $W$ provided the ansatz is sufficiently expressive. 

The ansatz circuit we propose is following a layered structure, where each layer $l$ consists of the product of an entangling part followed by single-qubit rotations, $U^s_l(\vec{\theta}_l^s)U^e_l(\vec{\theta}_l^e)$, where
\begin{align}
  \begin{split}
  U_l^s(\vec{\theta}_l^s)&=\prod_{k=0}^{NF-1}e^{-\frac{i}{2}(\vec{\theta}_l^s)_k\sz_k}, \\  U_l^e(\vec{\theta}_l^e)&=\prod_{k \text{ odd}}e^{-\frac{i}{2}(\vec{\theta}_l^e)_k(\sx_k\sx_{k+1}+\sy_k\sy_{k+1})}\prod_{k \text{ even}}e^{-\frac{i}{2}(\vec{\theta}_l^e)_k(\sx_k\sx_{k+1}+\sy_k\sy_{k+1})}.
  \end{split}
  \label{eq:QuAlg}
\end{align}
As one can easily show, the ansatz conserves the total charge. Moreover, for the specific choice $\nu_0=-\nu_{2}$, $\nu_1=0$, and $\mu_f=\text{const}.$ for all flavors, the symmetry under simultaneously flipping the spins and reflecting around the center can be ensured by choosing the parameters as $ (\vec{\theta}_l^e)_k = (\vec{\theta}_l^e)_{3N-2-k}$ for the entangling part and $(\vec{\theta}_l^s)_k = -(\vec{\theta}_l^s)_{3N-1-k}$ for the single-qubit rotations.

In addition, the initial state has to be chosen in the correct symmetry sector. A N\'eel state in the spin sites $\ket{\psi_0} = \ket{1010\dots}$ is both in the sector of vanishing total charge and symmetric under reflecting and flipping all spins around the center. Thus, the ansatz we propose is given by
\begin{align}
  \ket{\psi(\vec{\theta}^s_{L-1}, \vec{\theta}^e_{L-1}, \dots, \vec{\theta}^s_0,\vec{\theta}^e_0)} = \prod_{l=0}^{L-1}U_l^s(\vec{\theta}_l^s)U_l^e(\vec{\theta}_l^e)\ket{\psi_0}.
  \label{eq:ansatz_circuit}
\end{align}
For the general case, where the parameters inside each layer are not constrained, the ansatz has $p=6N-1$ parameters per layer. In case the symmetry constraint is enforced, this number can be reduced to $p=3N$. Note that the ansatz in Eq.~\eqref{eq:ansatz_circuit} can be readily implemented on measurement-based quantum computers as a series of measurements on an initially prepared cluster state~\cite{Schuster2022}.

\section{Results\label{sec:results}}
In order to benchmark the performance of the ansatz, we classically simulate the VQE assuming a perfect quantum computer without shot noise. The classical optimization procedure is carried out using the L-BFGS algorithm~\cite{Nocedal2006}. For each choice of chemical potential, we run 10 different simulations with randomly chosen initial parameters. We deem a result as an outlier if the energy obtained is more than 30\% higher than the lowest one observed during the 10 runs or the particle number is not an integer value to good approximation. In the following, we examine three parameter regimes: (i) vanishing bare fermion mass for which analytical results are available, (ii) non-vanishing bare fermion mass, and (iii) a sign-problem afflicted regime for conventional MC simulations.

\subsection{Vanishing bare fermion mass}
\begin{figure}[htp!]
  \centering
  \includegraphics[width=1.0\textwidth]{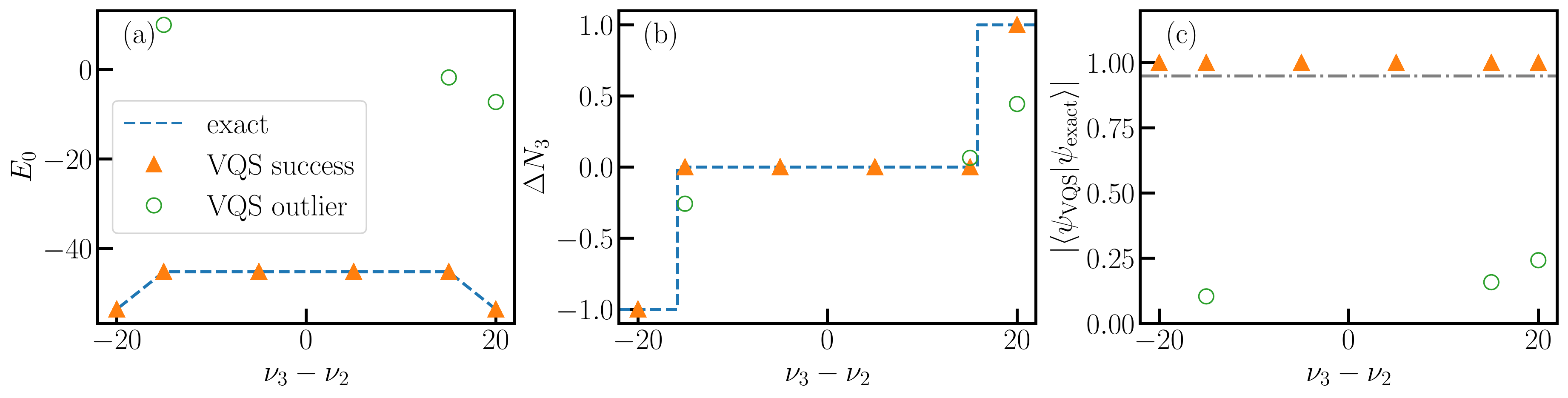}
  \includegraphics[width=1.0\textwidth]{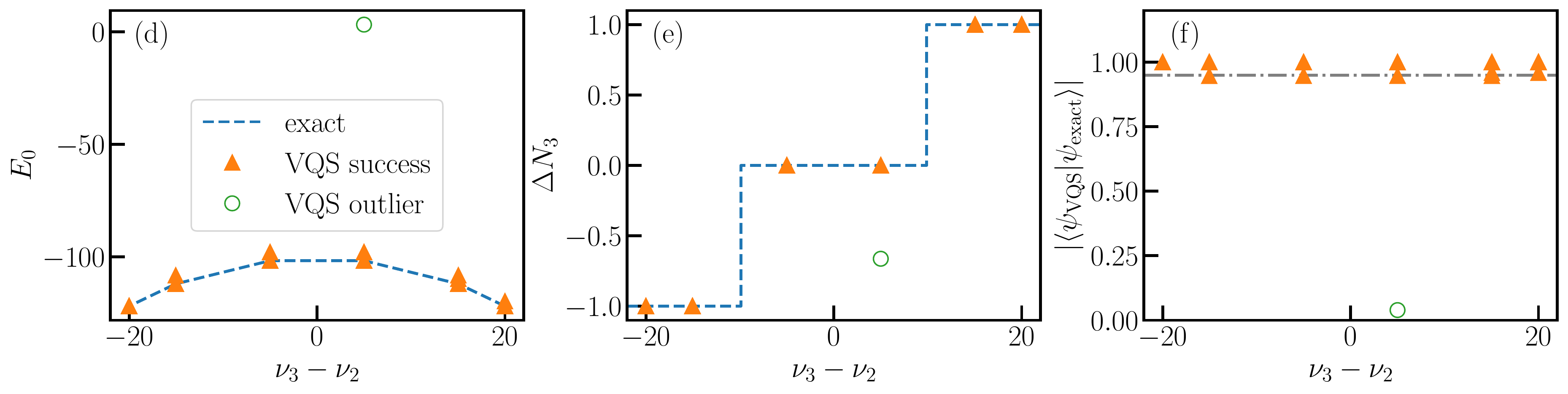}
  \includegraphics[width=1.0\textwidth]{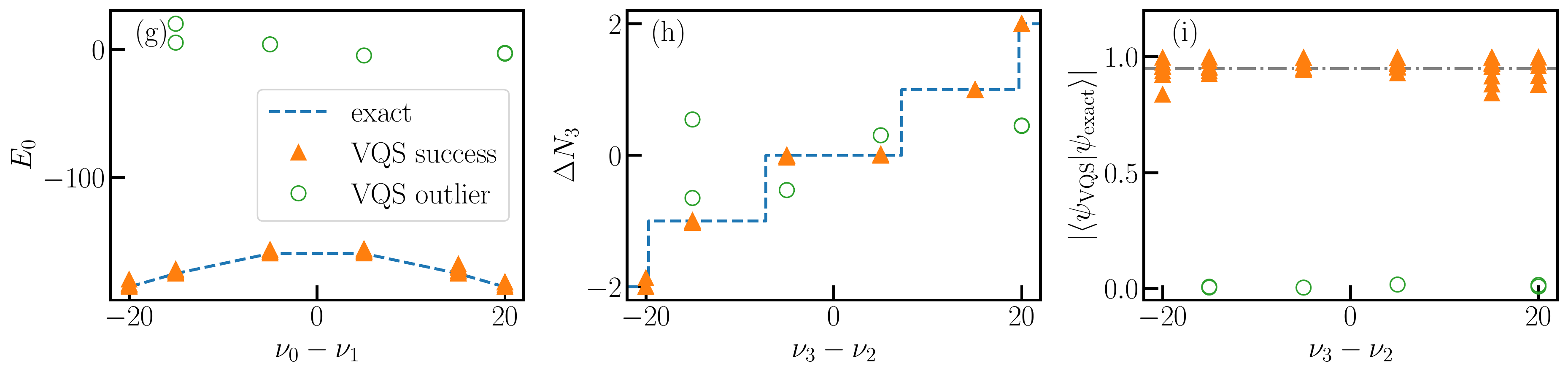}
  \caption{Ground-state energy (first column), particle number (second column), and overlap (third column) of the VQE solution as a function of the difference of the chemical potentials for $N=2$ (first row), $4$ (second row), $6$ (third row), and 5 layers of the ansatz. Successful runs correspond to filled orange triangles, outliers to open green circles. The dashed blues lines indicate the exact solution obtained via exact diagonalization. The dash-dotted gray line in panels (c), (f), and (i) indicates the 95\% threshold for the overlap.}
  \label{fig:N2m0}
\end{figure}

First, we focus on the case $\mu_f=0$, for which analytical results are available~\cite{Lohmayer2013}. Since the physics of the model only depends on the difference of the chemical potentials, we set $\nu_1=0$ without loss of generality. Moreover, we restrict ourselves to the case $\nu_2 = -\nu_0$, for which the model has the aforementioned reflection symmetry. The results of the simulations for this parameter set for system sizes $N=2$, $4$, and $6$, corresponding to $6$, $12$, and $18$ qubits, are shown in Fig.~\ref{fig:N2m0}.
In general, we observe good agreement between our VQE results and the exact solution obtained by exact diagonalization. In particular, our VQE runs are able to generate overlaps with the exact solution of more than 95\% for most cases. Despite keeping the number of layers constant for various system sizes, the overlaps only slightly decrease with increasing system size. Looking at the particle numbers in Fig.~\ref{fig:N2m0}, we observe the characteristic discontinuities indicating first-order quantum phase transitions between different phases, which are well captured by the VQE results.

Focusing on the outliers in Fig.~\ref{fig:N2m0}, these can be easily identified just by looking at the energy and the particle numbers, observables which can be measured efficiently. In particular, we see that the outliers have high energies and particle numbers that are non-integer, which clearly indicates that they are unphysical. The overlap confirms this observation, as the outliers correspond to solutions which have almost vanishing overlap with the exact one (cf. Fig.~\ref{fig:N2m0}).

All in all, the ansatz allows for capturing the physics of the model in this parameter regime and is able to resolve the first-order phase transitions expected from the theoretical prediction. Moreover, the success probability for the VQE is high and occasional outliers can be identified easily from the physical observables.

\subsection{Non-vanishing bare fermion mass}
\begin{figure}[htp!]
  \centering
  \includegraphics[width=1.0\textwidth]{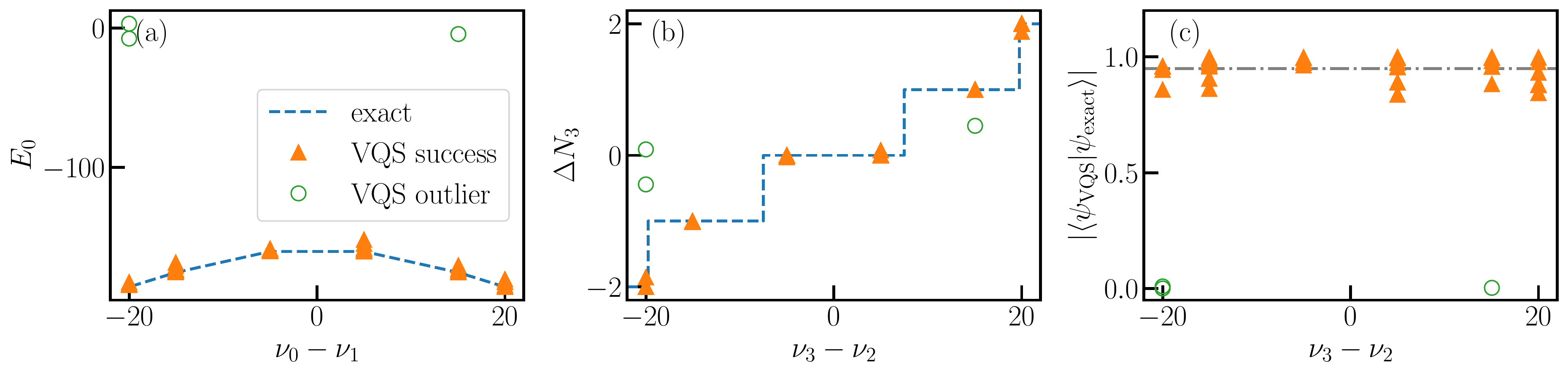}
  \caption{Ground-state energy (a), particle number (b), and overlap (c) of the VQE solution as a function of the difference of the chemical potentials for $N=6$, $\mu_f=0.1$, and 5 layers of the ansatz. Successful runs correspond to filled orange triangles, outliers to open green circles. The dashed blues lines in panels (a) and (b) indicate the exact solution obtained via exact diagonalization. The dash-dotted gray line in panel (c) indicates the 95\% threshold for the overlap.}
  \label{fig:N6m01}
\end{figure}

Second, we look at $\mu_f\neq0$, still keeping $\nu_1=0$ and $\nu_2 = -\nu_0$. While for this case there are no analytical predictions available, this parameter regime is in principle accessible with conventional MC methods. Our VQE results for 18 qubits corresponding to $N=6$ are depicted in Fig.~\ref{fig:N6m01}.
Compared to the case of vanishing fermion mass, we qualitatively observe the same behavior. The results for the energy and the particle number are in good agreement with the exact solution, and a large fraction of our runs is able to produce overlaps with the exact ground state of 95\% and above. Furthermore, outliers can be easily identified from the physical observables.

\subsection{Sign-problem afflicted regime}
\begin{figure}[htp!]
  \centering
  \includegraphics[width=1.0\textwidth]{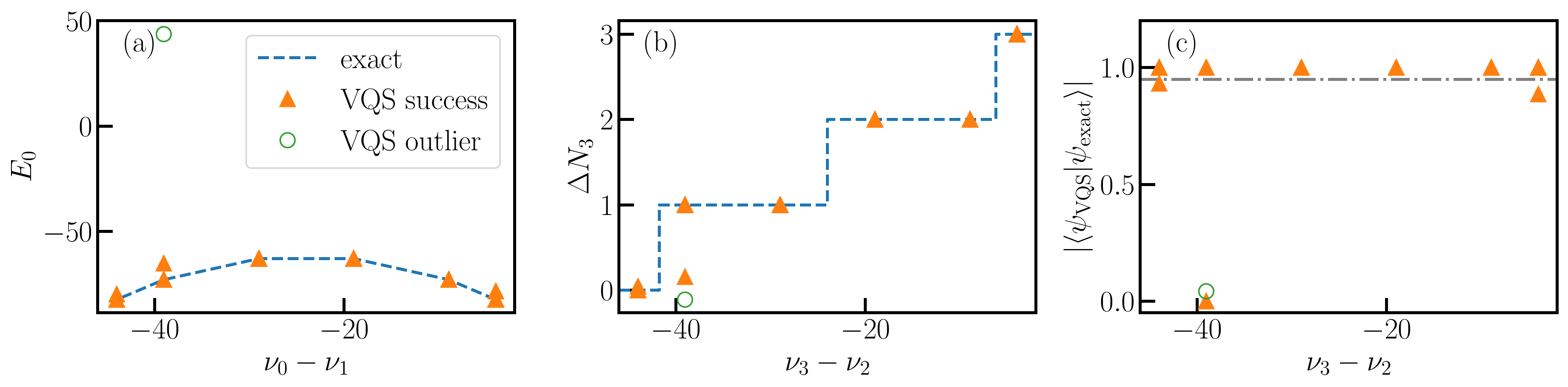}
  \caption{Ground-state energy (a), particle number (b), and overlap (c) of the VQE solution as a function of the difference of the chemical potentials for $N=4$, $\mu_f=0$, and 5 layers of the ansatz. Successful runs correspond to filled orange triangles, outliers to open green circles. The dashed blues lines in panels (a) and (b) indicate the exact solution obtained via exact diagonalization. The dash-dotted gray line in panel (c) indicates the 95\% threshold for the overlap.}
  \label{fig:N4m0_kg3}
\end{figure}

Finally, we look at $\mu_f=0$ and $\nu_2=-\nu_0$, but this time with $\nu_1=3.0$, a regime which is inaccessible with conventional MC methods due to the sign problem. Note that, for this case, the reflection symmetry around the center followed by a spin flip no longer holds true and we cannot constrain some of the parameters in the ansatz anymore. The results of the simulations for this parameter regime are shown in Fig.~\ref{fig:N4m0_kg3}.
%
%
Even for this situation, where conventional MC approaches fail, our VQE ansatz manages to produce results that are in good agreement with the exact solution. Despite the fact that we can no longer constrain the ansatz using the symmetry and for $N=4$ we now have we have 23 parameters per layer instead of 12 previously, the optimization still works reliably and we do not see a substantial increase in the number of cases where the VQE fails, as a comparison between Figs.~\ref{fig:N2m0}(f) and \ref{fig:N4m0_kg3}(c) reveals. We only observe a single data point for which the energy of the VQE result is close to the exact one, the particle number is approximately an integer value, but the overlap with the exact solution vanishes. This happens nearby the phase-transition point, at which there exist two degenerate energy levels in the Hamiltonian. In the vicinity of this point, the energy levels are still close and the VQE likely converged to the wrong one of them, hence explaining why the energy is similar to the exact solution but the particle number differs and the overlap vanishes. In general, despite the increased number of parameters, the ansatz still shows good performance and produces high overlaps with the exact solution for almost all runs, even in the sign-problem afflicted regime.

\section{Conclusion and outlook\label{sec:conclusion}}
In summary, we have proposed an ansatz circuit for a VQE solving the lattice Schwinger model with three fermion flavors in the presence of a chemical potential. The ansatz allows for incorporating the relevant symmetries of the model and can be implemented on both circuit-based and measurement-based quantum devices.

Simulating the VQE classically by assuming a perfect quantum computer without shot noise, we have benchmarked the performance of the ansatz and demonstrated that it allows for obtaining a good approximation of the ground state of the model, even in regimes where MC methods suffer from the sign problem. In particular, we have shown that our ansatz is able to resolve the first-order phase transitions that occur in the model. Moreover, our results for various system sizes indicate that the number of layers required to capture the relevant physics does not grow strongly with the number of lattice sites.

In the future, we aim to investigate the performance of the ansatz in the presence of noise and to implement a proof-of-principle simulation on a noisy, intermediate-scale quantum device.

\acknowledgments
L.F.\ is partially supported by the U.S.\ Department of Energy, Office of Science, National Quantum Information Science Research Centers, Co-design Center for Quantum Advantage (C$^2$QA) under contract number DE-SC0012704, by the DOE QuantiSED Consortium under subcontract number 675352, by the National Science Foundation under Cooperative Agreement PHY-2019786 (The NSF AI Institute for Artificial Intelligence and Fundamental Interactions, \url{http://iaifi.org/}), and by the U.S.\ Department of Energy, Office of Science, Office of Nuclear Physics under grant contract numbers DE-SC0011090 and DE-SC0021006.
S.K.\ acknowledges financial support from the Cyprus Research and Innovation Foundation under projects ``Future-proofing Scientific Applications for the Supercomputers of Tomorrow (FAST)'', contract no.\ COMPLEMENTARY/0916/0048, and “Quantum Computing for Lattice Gauge Theories (QC4LGT)”, contract no.\ EXCELLENCE/0421/0019.
This work was funded by the Deutsche Forschungsgemeinschaft (DFG, German Research Foundation) -- Project-ID 429529648 -- TRR 306 QuCoLiMa (``Quantum Cooperativity of Light and Matter''). 

\bibliographystyle{JHEP}
\bibliography{Papers}
\end{document}